\documentclass[%
 reprint,
 amsmath,amssymb,
 aps,
longbibliography,
showkeys]{revtex4-2}

\usepackage{graphicx}
\usepackage{dcolumn}
\usepackage{bm}
\usepackage{soul}

\makeatletter
\def\fourvdots{\vbox{\baselineskip1\p@ \lineskiplimit\z@\kern6\p@\hbox{.}\hbox{.}\hbox{.}\hbox{.}}}
\makeatother
\usepackage{hyperref}

\usepackage{physics}
\usepackage{bbm}
\usepackage[dvipsnames]{xcolor}

\begin{document}

\preprint{APS/123-QED}

\title{Thermally Robust Cooperative Function in Mechanical Networks}

\author{Ben Pisanty$^{1}$}
\author{Jovana A. Kim$^{1}$}%
\author{Andrea J. Liu$^{1,2}$}
\author{Sidney R. Nagel$^{3}$}
\affiliation{$^1$Department of Physics and Astronomy, University of Pennsylvania, Philadelphia, Pennsylvania 19104, USA}%
\affiliation{$^2$Santa Fe Institute, Santa Fe, NM 87501}
\affiliation{$^3$Department of Physics and The James Franck and Enrico Fermi Institutes, The University of Chicago, Chicago, Illinois 60637, USA}

\date{\today}

\begin{abstract}
Elastic networks can be tuned to exhibit complex mechanical responses and have been extensively used to study protein allostery, where a localized strain regulates the conformation at a distant site. 
We show that cooperative binding, where one site enhances the other's ability to function, can be improved by tuning for \textit{bidirectional} allostery: a symmetric coupling where the strain propagated between the two sites is independent of which site is perturbed.
We identify a crossover temperature above which functionality breaks down due to thermal fluctuations.
We introduce a modified tuning process to increase this crossover temperature, showing that function can be robust at biologically relevant temperatures if thermal stability is important.
\end{abstract}


\maketitle

\section{\label{sec:level1}Introduction}
Many proteins exhibit allostery, where binding at a source, known as the allosteric or regulatory site, affects binding at a target, 
the active site, to regulate biological processes ~\cite{MONOD196588, Changeux2011, Tsai2014Plos, WODAK2019566}. 
Folded proteins are often modeled as bead-spring networks~\cite{Tirion1996,Bahar1997A,Bahar1997B, Hinsen1998, PAK2018Advances, Flechsig2019Simple}. When viewed through this mechanical lens~\cite{Mitchell2016Strain, Flechsig2017Design, Flechsig2019Simple}, allostery occurs when binding-induced strain at the allosteric site (source) is transmitted through the protein to produce specified strain at the active site (target) that enables or prohibits binding. 
One approach to understanding allostery has been to design mechanical spring networks that produce a localized target strain in response to a localized strain at the source~\cite{Flechsig2017Design, Rocks2017Designing, Yan2017PNAS,Yan2018Principles, WODAK2019566, Ravasio2019Mechanics, Rocks2019Limits, Bravi2020Plos, Rouviere2023Emergence, rocks2024topological}.

Cooperative binding is an important function of allosteric mechanisms that measures cross-communication between multiple binding sites ~\cite{RUBIN1966_On_the_nature, Stefan2013Cooperative_Binding}. 
It is particularly important when considering proteins with equivalent binding sites, such as hemoglobin. In its simplest form, the mechanism for such proteins can be modeled via bidirectional allostery; a source strain at one location produces a target strain at the other location and vice versa. Mechanical networks can display allosteric relations between multiple source-target pairs \cite{Rocks2019Limits}, but bidirectional allostery requires the response to be symmetric. We adapt an existing process for tuning allostery to optimize for bidirectional allostery instead. 
As in previous work~\cite{Flechsig2017Design, Rocks2017Designing, Yan2017PNAS,Yan2018Principles, Ravasio2019Mechanics, Rocks2019Limits, Bravi2020Plos, Rouviere2023Emergence, rocks2024topological}, we study ensembles of networks tuned at zero temperature.

A mechanical network that exhibits allostery has displacements due to the source strain that extend across the system to the target. These displacements tend to project strongly onto a low-frequency vibrational normal mode~\cite{Yan2017PNAS,Yan2018Principles, Ravasio2019Mechanics,Loutchko2020Allosteric}. 
This result is quite general; tuning a physical network with reciprocal interactions to develop desired responses leads to low-lying eigenvalues of the physical Hessian ~\cite{Stern2024Physical_effects_of_learning, stern2024Physical_Networks}. Various algorithms have been used to tune mechanical networks for function, including discrete gradient descent \cite{Goodrich2015Tuning, Rocks2017Designing}, directed aging \cite{Pashine2019Directed}, fitness Monte Carlo \cite{Yan2017PNAS,Yan2018Principles}, and evolutionary optimization \cite{Flechsig2017Design, Tlusty2017genotype}.
Despite their differing tuning processes, the resulting mechanical networks share a common property: the emergence of a low-frequency  mode that dominates the network’s functional response.
Thus, low-lying eigenmodes, which correspond to low-frequency vibrational modes in mechanical networks, are a generic result of tuning, independent of the tuning process or function \cite{Guzman2025Imprints}. Indeed, such modes have been reported and associated with function in allosteric \cite{Zheng2006, Glembo2012, Townsend2015, Eckmann2019Colloquium} and other proteins \cite{Husain2020Epistasis,Russo2025Soft} through computation and simulation, and the slow, large-amplitude collective motions they describe have been directly observed experimentally in allosteric proteins via NMR relaxation methods~\cite{HenzlerWildman2007}. However, proteins function at a nonzero temperature, and such a low-frequency response will inevitably be sensitive to thermal excitations that can alter or destroy the function.
This poses questions about the stability of function to fluctuations: Can fluctuations lead to functional failure? Are there tradeoffs between function and thermal stability?

Here we explore how thermal fluctuations affect tuned mechanical networks. Specifically, we use analytical and numerical methods to study the effects of temperature on bidirectional allostery, identifying a crossover temperature, $T^*$, above which the functionality is destroyed. Based on estimates matching our networks' dimensions and stiffness to those of proteins, we find that proteins operate in a thermally challenging environment, near $T^*$. 
This suggests that robustness to fluctuations caused by the environmental temperature is an important constraint that must be satisfied by protein evolution. Here, we introduce a modified tuning protocol that raises $T^*$ so that functionality is more robust to higher temperatures. To do this, we not only tune for function but also simultaneously for thermal stability.

\section{Tuning Cooperative Function via bidirectional allostery}\label{Network Tuning}
We generate an ensemble of three-dimensional disordered central-force spring networks, and tune them at zero temperature to achieve the desired functionality \cite{Kim_physical_learning_2026}. It is now generally accepted that in allosteric regulation, ligand binding changes the energy landscape of the protein \cite{Gunasekaran2004,Smock2009}, altering the structure or dynamics at a distant site. In our model, binding is represented by the addition of a pre-stressed spring at the source site, causing the system to favor a new equilibrium configuration. In this simple mechanical model, strain represents the conformation coordinate. We define allosteric functionality of magnitude $\Delta$ as a strain $\varepsilon_t=\varepsilon_s \Delta$ at a target site in response to a strain $\varepsilon_s$ at a source site~\cite{Rocks2017Designing, Rocks2019Limits}. Under this definition, $\Delta$ represents a partial conformation change in the unoccupied binding site towards the bound conformation \cite{Kern2003}. As will be discussed later, thermal fluctuations allow the network to transition between the unbound and bound conformations, consistent with the ensemble view of proteins \cite{Motlagh2014}.

To increase \textit{cooperative} function, we tune for bidirectional allostery, where we require allosteric functionality of magnitude $\Delta$ at target site B in response to strain applied at source site A and simultaneously allosteric functionality of the same magnitude $\Delta$ at A in response to strain applied at B.

\begin{figure}
\includegraphics[width=.99\linewidth]{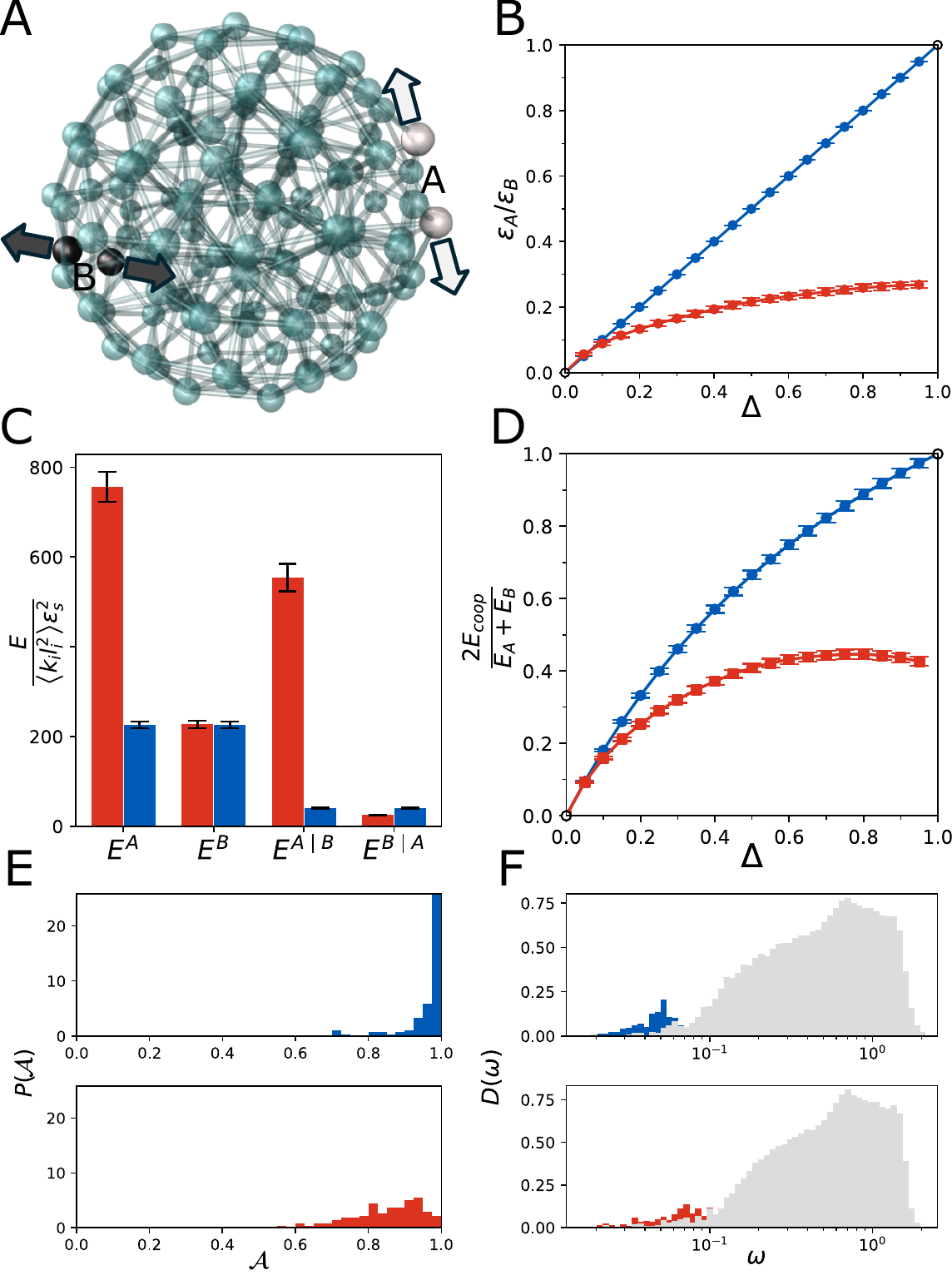}
\centering
\caption{\textbf{One-directional and Symmetric tuning.} 
(A) A randomly generated spring network; a task $A\rightarrow B$ ($B\rightarrow A$) with allosteric ratio $\Delta$ is defined by the black (white) nodes moving apart and achieving a target strain $\varepsilon_B$ ($\varepsilon_A$) in response to a source strain $\varepsilon_A$ ($\varepsilon_B$) imposed on the white (black) nodes, such that $\varepsilon_B/\varepsilon_A = \Delta$ ($\varepsilon_A/\varepsilon_B = \Delta$). 
(B) The reverse response ($B\rightarrow A$) of networks tuned for $A\rightarrow B$ (red squares), and the response ($A\rightarrow B$ or $B\rightarrow A$) of networks tuned symmetrically (blue circles).
One-directional tuning results in a weak response when the sites of source and target are reversed. Open symbols at $\Delta=0,1$ are determined by theoretical arguments, not simulation. 
(B-F) show aggregated data, error bars in (B-D) indicate standard errors.
(C) Binding energies of networks tuned for $\Delta=0.7$; $E^A$ and $E^B$ are the free state energies required to strain one of the sites; $E^{A|B}$ and $E^{B|A}$ are the additional energies required to strain the second site. For one-directional tuning, the energy barrier $E^{A|B}$ is larger than $E^B$. The vertical axis is scaled by the typical energy required to strain a bond, $\langle k_i l_i^2\rangle \varepsilon_s^2$.
(D) Cooperativity; symmetric tuning results in increased cooperation. The curves agree with the analytically calculated data points at $\Delta=0,1$. 
(E) Identifying the cooperative/allosteric mode involves decomposing the free state response using the normal modes, and singling out the mode with the highest projection coefficient. Here we plot the maximal projection coefficient $\mathcal{A}$ for systems tuned for $\Delta=0.7$, showing that the free state is highly aligned with the cooperative/allosteric mode.
(F) Density of states for systems tuned for $\Delta=0.7$. Colored bars indicate the frequency of the cooperative/allosteric mode (one for each system), showing that it lies at low frequencies compared to the rest of the modes (gray). Frequencies $\omega$ were derived from the eigenvalues of each system, which were first normalized with respect to the average stiffness. 
}
\label{fig1}
\end{figure}

We construct over 100 three-dimensional networks from the contact network of randomly generated packings of 128 bi-disperse soft spheres (radius ratio of 0.8 between the large and small spheres, with each sphere having an equal probability of being large or small). The spheres are enclosed in a soft external shell bounded by the unit cube and compacted to a mean contact number of $z = 7.4\pm 0.2$ by shrinking the shell in a damped molecular dynamics simulation. Nodes 
of mass $m=1$ 
are placed at the center of each sphere, with average center--center distance of $ l_0 =0.086$. The unstretched spring stiffnesses are initialized to $k_0 = 10^{-2} k_{\text{max}}$, where the maximum stiffness is defined as $k_{\text{max}} = 1$, yielding $k_0 = 0.01$. Underconstrained nodes on the surface are removed until the network has no floppy modes \cite{Rocks2017Designing}. We randomly select two pairs of neighboring nodes on opposing sides of the surface of the packing to serve as the two ``binding" sites A and B (see Fig. \ref{fig1}A), and following Ref.~\cite{Rocks2017Designing}, remove the edge connecting each of these pairs.

To achieve cooperative functionality, we tune the spring stiffnesses, $k_i$ of each bond (edge) $i$ in the network using the coupled learning algorithm \cite{Stern2021Supervised}. 
This tuning process defines two states with different boundary conditions. In the \textit{free state}, an external strain $\varepsilon_s$ is applied only between the pair of nodes at the first (source) site, inducing a response $\varepsilon_t$ at the second (target) site. In the \textit{clamped state} the source strain $\varepsilon_s$ is applied as before, while an additional strain $\varepsilon_t+\eta(\varepsilon_s \Delta - \varepsilon_t)$, with $\eta \ll 1$, is imposed at the target site, thereby gently nudging its response towards the desired strain $\varepsilon_s \Delta$.

In this contrastive learning algorithm, the spring stiffnesses, which mediate interactions between the nodes, serve as \textit{tunable} degrees of freedom. These degrees of freedom are adjusted to minimize the contrast function, 
\begin{equation}\label{eq1: contrast}
\mathcal {C} \equiv E^C-E^F,
\end{equation} 
where, $E^C$ and $E^F$ are the elastic energies of the clamped and free states, derived from spring stiffnesses and the \textit{physical} degrees of freedom, namely the node positions.
This minimizes the difference between the free and clamped states. 
To accomplish this, we tune spring stiffnesses in conjunction with a $T=0$ overdamped molecular dynamics simulation which governs the node positions. 

To tune cooperative functionality, we use a symmetric application of coupled learning, where two sites (A and B) serve as both source and target to one another. We simultaneously consider four copies of the system, each subjected to a different set of strain boundary conditions at sites A and B. The first two copies correspond to the free and clamped states when the input strain is applied at A (denoted by $A \rightarrow B$). The third and fourth copies correspond to the free and clamped states when the input strain is applied at B ($B \rightarrow A$). The tunable degrees of freedom ($k_i$) are the same across all four copies. The physical degrees of freedom (node positions) are adjusted separately for each copy to minimize the energy under the applied boundary condition.
The spring constants are tuned using: 
\begin{equation}\label{eq1: symmetric learning rule}
\frac{dk_i}{dt}=-\frac{\alpha}{\eta}\frac{d}{d k_i}\left( \mathcal {C}_{A\rightarrow B} + \mathcal {C}_{B\rightarrow A}\right),
\end{equation}
where $\alpha$ is the learning rate, and $\eta$ is the nudge factor introduced earlier. 
Two time scales govern the dynamics of our tuning process. The mechanical relaxation time, $\tau_m\sim2m/b$, is set by the mass of the nodes and the viscous damping coefficient $b$. The adaptation time, $\tau_k\sim k/\left( \alpha \eta \varepsilon^2 l^2 \right)$, on the other hand, is set by the tuning and network's parameters. In our simulations, the mechanical relaxation time is much smaller than the adaptation time, $\tau_m\ll\tau_k$. Although we work near the quasi-static limit, we note that is not necessary for the node positions to equilibrate fully before each adjustment of the tunable degrees of freedom, and that the tuning process avoids oscillatory learning dynamics up to $\tau_m\sim\tau_k$ ~\cite{Stern2022Physical}.

Using Eq.~\ref{eq1: contrast} and carrying out the derivatives in Eq.~\ref{eq1: symmetric learning rule}, we obtain the local learning rule:
\begin{multline}\label{local learning rule}
\frac{dk_i}{dt}=\frac{\alpha}{\eta}\biggl[ \frac{1}{2}\left( \left({\Delta l_i^F}\right)^2 -\left({\Delta l_i^C}\right)^2 \right)_{A\rightarrow B} \\
+  \frac{1}{2}\left( \left({\Delta l_i^F}\right)^2 -\left({\Delta l_i^C}\right)^2 \right)_{B\rightarrow A} \biggr]
\end{multline}
where $\Delta l_i$ is the elongation of spring $i$ with respect to its rest length $l_i$, and $\langle l_i \rangle=l_0=0.086$.
Before tuning, all springs have a uniform initial stiffness $k_0=0.01$ and are at their rest lengths in the absence of any applied strain. The minimum and maximum allowed stiffnesses are $10^{-2}k_0$, (essentially no bond, as in \cite{Rocks2017Designing,Yan2017PNAS,Bravi2020Plos}) and $10^2k_0$.
This range is consistent with coarse-grained elastic models of proteins~\cite{Smith2007Bio, Voth2008Bio, Robert2009Pnas,Ming2005PRL,McLeish2022Jour}.

In Fig.~\ref{fig1}B, the blue circles show that for symmetric tuning, the system achieves bidirectional allostery; the response strain of site A to input strain at site B and of site B to strain at site A are both close to the desired value $\Delta$. The red squares show that when the system is trained for uni-directional response $A\rightarrow B$, the response in the opposite direction $B \rightarrow A$ is much smaller.

Cooperation between binding sites is the degree to which binding at one site facilitates (or restricts) binding at the other. We use the elastic energy corresponding to the energy cost of a specified strain, as a proxy for the binding energy barrier. 
To quantify the degree of cooperativity, we calculate the cooperative energy $E_{coop} = E^A+E^B-E^{AB}$ \cite{Yan2018Principles, Zhu2011}, which can be written as:
\begin{equation}\label{eq2.1: coop energy partial 1}
E_{coop} = \frac{1}{2}\left( \left( E^A - E^{A|B} \right) + \left( E^B - E^{B|A} \right) \right),
\end{equation}
where $E^A$ and $E^B$ are the energies required to strain each site individually, and $E^{A|B}$ ($E^{B|A}$) is the energy required to strain site A (B) given that site B (A) is already strained. For cooperative binding, we want $E^{A|B}< E^A$ and $E^{B|A}<E_B$. That is, it should cost less energy to strain each site when the other is already strained than when it is not. Note that Ref~\cite{Yan2018Principles} tuned networks to raise $E_{coop}$ directly instead of tuning bidirectional strain tasks.

Figure~\ref{fig1}C compares $E^A$, $E^B$, $E^{A|B}$ and $E^{B|A}$ for one-directional allostery tuning (red) and bi-directional symmetric tuning (blue) at $\Delta=0.7$.
It is useful to define the nondimensional cooperativity parameter $R$:
\begin{equation}\label{eq3: cooperativity}
R = \frac{2E_{coop}}{E^A + E^B}.
\end{equation}
In the absence of any cooperativity, straining one site does not affect the energy required to strain the other site and $R=0$. If it costs no additional energy to strain the other site ($E^{B|A}=0$ and $E^{A|B}=0$) then the system is fully cooperative and $R=1$. The cooperativity parameter for allosteric unidirectional tuning (red) versus symmetric bidirectional tuning (blue) is shown in Fig.~\ref{fig1}D. 
Fig.~\ref{fig1}D shows that symmetric tuning leads to higher cooperativity $R$.

In the absence of tuning, the target site can either elongate or compress as a result of a strain at the source; the energy required to apply a strain in a desired direction is equally likely to increase or decrease.  Therefore on average $\Delta=0$ and $R=0$, consistent with Fig.~\ref{fig1}D.

Tuning the network to $\Delta>0$ leads to the development of a low-frequency mode that is aligned with the functionality \cite{Yan2018Principles,stern2024Physical_Networks}.
As we tune the network progressively to higher values of $\Delta$, the frequency of the mode decreases and the free state projects more strongly onto the mode. 
Figures~\ref{fig1}E-F show that for an ensemble of networks at $\Delta=0.7$ the free state already projects primarily on one mode, which typically is also the mode with the lowest frequency. On average, the projection onto the relevant mode is 0.84 for uni-directional tuning and 0.95 for bi-directional tuning. For the case of bi-directional (one-directional) tuning, we will refer to this mode as the ``cooperative mode" (``allosteric mode").

We find that successfully tuned bidirectional cooperative tasks fall in the range $0 \le \hat{\Delta} \equiv \varepsilon_t/\varepsilon_s \le 1$, where $\hat{\Delta}$ is the attained response. 
Note that this behavior is quite different from uni-directional allostery, where arbitrarily high values of $\hat{\Delta}$ can be tuned in the linear regime~\cite{Rocks2017Designing}.
In the limit $\hat{\Delta}=1$, $E^{B|A}$ and $E^{A|B}$ both vanish, so the system is fully cooperative ($R=1$). 
Fig. \ref{fig1}D shows that the cooperativity indeed approaches $R=1$ as $\Delta \rightarrow 1$. 
In that limit, we find that the displacements collapse onto a single cooperative mode that becomes floppy (its frequency vanishes). For uni-directional allostery, by contrast, we find that displacements collapse onto a single floppy mode in the limit $\Delta \rightarrow \infty$. Note that it should be possible to find balanced force states at $\Delta=1$ that are not floppy, but our tuning process does not find such states.

\section{Crossover Temperature}\label{crossover temp}
The tuning process presented in Sec.~\ref{Network Tuning} is designed to enhance cooperative function via bidirectional allostery at zero temperature. It means that the strain at the responding site approaches $\varepsilon_s\Delta$, where $\varepsilon_s$ is the strain applied to the other site. In a thermal environment, however, the network experiences fluctuating forces that strain all the springs.
With increasing temperature, the fluctuating strain at one site is increasingly likely to reach the desired value even in the absence of externally applied strain at the other site. If that occurs, then binding at the target does not require binding at the source, so the functionality is destroyed.

\begin{figure}
\includegraphics[width=.99\linewidth]{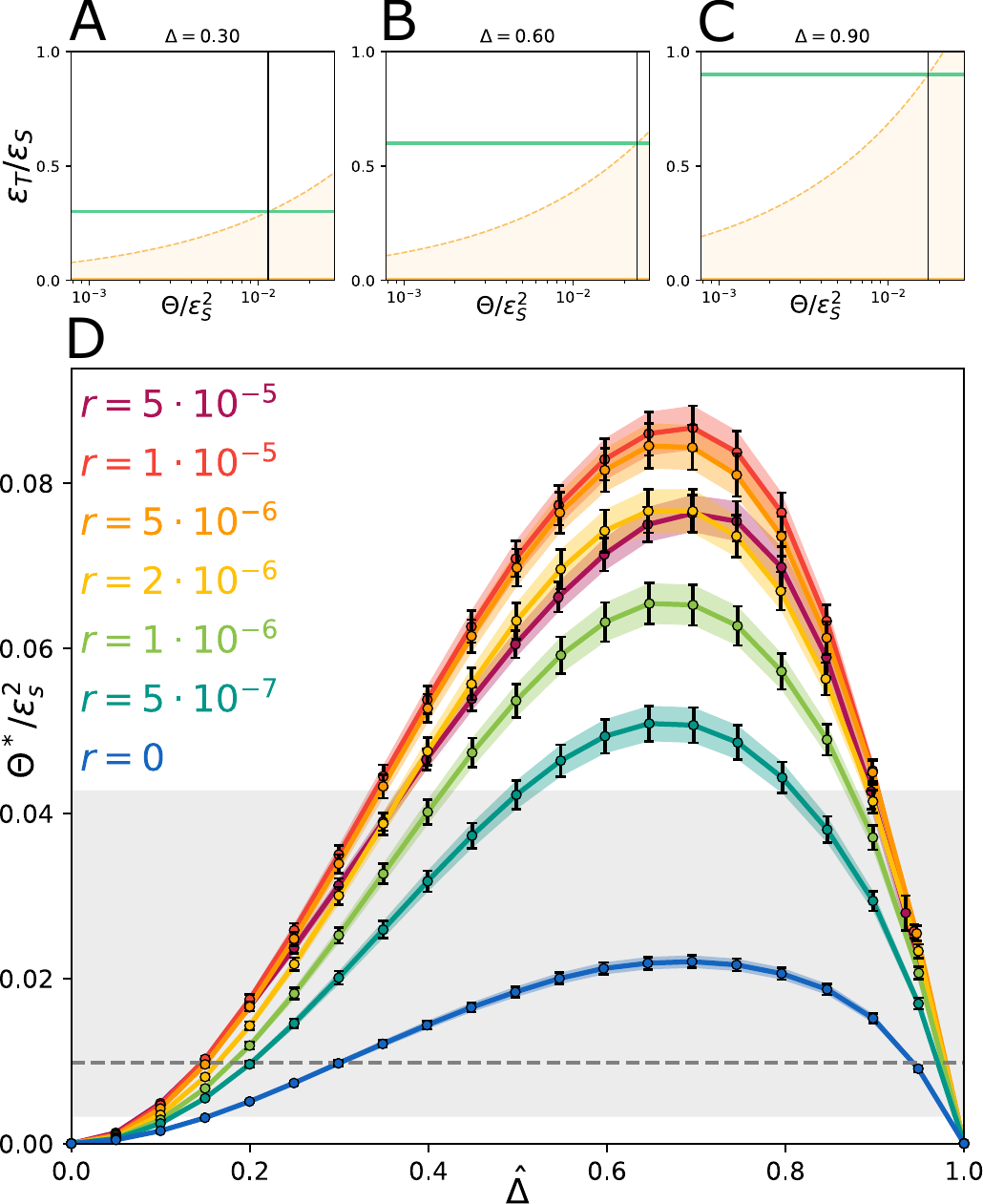}
\centering
\caption{\textbf{Crossover Temperature.} 
(A-C) Analytical calculations of the crossover temperature of a particular network tuned for increasing values of $\Delta$. The solid and dashed lines depict the expected and 1-sigma fluctuations range of the target strain for the cases with (green) and without (yellow) binding at the source. The temperatures are scaled with respect to the characteristic energy scale of the springs $\Theta=K_BT/\langle k_i l_i^2\rangle$ and to the input binding strain $\varepsilon_s$. The target strain is also scaled with respect to the input strain and matches $\Delta$ (solid green line). The crossover temperature $\Theta^*$, where the green and yellow regions first intersect, is indicated with a vertical black line. As $\Delta$ is increased, the crossover temperature first increases and then decreases. This non-monotonic behavior is explored in the blue curve in the panel below. 
(D) The dimensionless crossover temperature $\Theta^*$ as a function of the attained response ratio $0.05\leq\hat{\Delta}\leq0.95$  at increasing values of the parameter $r$, the relative importance of raising the energy of the free state as defined in Eq.~\ref{eq: modified cost function}. $\Theta^*=0$ at $\hat{\Delta}$ because $\left<\varepsilon_t\right>=0$, and at $\hat{\Delta}$ because the systems become floppy.
Vertical and horizontal bars (too small to see) indicate standard errors.
The gray dashed line corresponds to $300\ \text{K}$, obtained by matching our coarse-grained model with typical (median) protein values. The shaded gray region reflects the 25th-75th percentile range across different proteins.
}
\label{fig2}
\end{figure}

Here, we calculate the strain fluctuations at the target as a function of the temperature in the linear response regime. We define the crossover temperature $T^*$ as when one-$\sigma$ thermal fluctuations at the target without the source strain (orange dashed line in Fig.~\ref{fig2}A-C), which we refer to as the ``unbound case", reach $\hat{\Delta}$, the attained target strain when the source is applied (solid green line), which we refer to as the ``bound case". This physical picture is conveyed for three different values of $\Delta$ in Fig.~\ref{fig2}A-C. For each value of $\Delta$ we show the mean strain at the target in the bound case (solid green) and the mean and standard deviation of the target strain in the unbound case (orange) as a function of the dimensionless temperature $\Theta=K_BT/\langle k_il_i^2\rangle$. The dimensionless temperature is scaled by $\varepsilon_s^2$ to offset the dependence of the crossover temperature on the input strain.

As derived in \textit{Appendix \ref{apB}}, the strain fluctuations in linear response at the network's bonds can be written as a sum of contributions from the system's normal modes. 
We are interested in the fluctuations at the two binding sites. Each of them can be considered as a target for a source strain applied at the other site. The standard deviation of strain fluctuations at the target can be written as
\begin{equation}\label{strain fluctuations}
\sigma^2_{\varepsilon_t} = \frac{1}{l^2_t}\sum_j\frac{k_BT}{m\omega^2_j} \left( \vec{R}_t \cdot \vec{e_j} \right)^2 
\end{equation}
where $l_t$ is the equilibrium length of the target bond, $\vec{R}_t$ is the row of the rigidity matrix $\mathbf{R}$ corresponding to the target site, $m\omega^2_j$ is the eigenvalue corresponding to eigenvector $\vec{e_j}$ of mode $j$, and the sum over eigenmodes excludes the trivial zero modes corresponding to global translations and rotations.

In \cite{supplemental}, we compare the strain fluctuations predicted by Eq.~\ref{strain fluctuations} to those observed in molecular dynamics simulations of the same networks (LAMMPS \cite{Gissinger24LAMMPS}). We find that the linear response calculation predicts the magnitude of strain fluctuations extremely well even for input strains in the range 0.1--0.3 ($10\%$--$30\%$), far beyond the linear regime~\cite{Sheinman2012Nonlinear, Rocks2017Designing}.  We also find that it works well up to surprisingly high temperatures, even two orders of magnitude above the crossover temperature that we define below. 

Specifically, we are interested in the fluctuations of the target strain in the absence of strain applied at the source, which we refer to as the unbound case. From Eq.~\ref{strain fluctuations} we see that it depends on temperature as
\begin{equation}
  \sigma^{UB}_{\varepsilon_t} = \sqrt{k_BT}\tilde{\sigma}^{UB}_{\varepsilon_t}  
\end{equation}
where $\tilde{\sigma}^{UB}_{\varepsilon_t}$ is independent of temperature. 

The mean strain at the target $\langle\varepsilon_t\rangle$ can also be derived from linear response. In the unbound case, the target strain fluctuates about zero so $\langle \varepsilon_t \rangle =0$. 
With an applied source strain, which we refer to as the bound case, it fluctuates about a non-zero value $\langle \varepsilon_t \rangle \approx \varepsilon_s \Delta$.

We now define a crossover temperature above which thermal fluctuations destroy bidirectional allostery. As mentioned earlier, we regard the unbound and bound cases as qualitatively distinct if the mean strain values for the two cases are more than the 1-sigma fluctuations in the unbound case. We therefore define the crossover temperature $T^*$ using 
\begin{equation}\label{crossover condition}
    \langle \varepsilon_t \rangle = \sqrt{k_BT^*} \left(\tilde{\sigma}^{UB}_{\varepsilon_t} \right).
\end{equation}

We note that this estimate does not include the effects of the fluctuations $\sigma^{B}_{\varepsilon_t}$ that are present when the source is bound, which are typically smaller than in the unbound case, see \cite{supplemental}. An alternative definition of the crossover temperature that also accounts for fluctuations in the bound state is discussed in \cite{supplemental} and yields qualitatively similar results.    

We turn now to the dependence of the dimensionless crossover temperature $\Theta^*$ on the amplitude of the tuned response, $\Delta$, plotted as the dark blue symbols and curve in Fig~\ref{fig2}D. For $\Delta=0$, there is no difference between the bound and unbound case, so clearly $\Theta^*=0$. At the other extreme, $\Delta=1$, the strain fluctuations at the target, $\sigma^{UB}_{\varepsilon_t}$, diverge at any non-vanishing temperature because the frequency of the cooperative mode vanishes (see Eq. \ref{strain fluctuations}). 
As a result, $\Theta^*=0$ at $\Delta=1$ and the behavior of $\Theta^*$ is nonmonotonic in $\Delta$. For small $\Delta$, the increase of $\langle \varepsilon_t \rangle \approx \varepsilon_s \Delta$ means that the responses at the target with and without the source strain are easier to distinguish, so $\Theta^*$ increases. For sufficiently large $\Delta$, however, the decrease of the frequency of the cooperative mode causes the fluctuations in the unbound case, $\sigma^{UB}_{\varepsilon_t}$, to increase, allowing them to reach $\varepsilon_t$ at lower temperatures, leading to a decrease in $\Theta^*$. 
The aggregated results for $\Theta^*$ from different systems are shown in the dark blue curve in Fig.~\ref{fig2}D.

\section{Dimensionless temperature estimations for proteins}

Converting from dimensionless temperature $\Theta$ calculated in our simulations to real temperature for real proteins requires matching the thermal energy scale $k_BT$ with the mechanical scale set by protein size and stiffness. To interpret the ratio $\Theta/\varepsilon_s^2$, we also need a typical strain that represents binding in proteins. 

The melting temperatures of proteins fall in a rather narrow range of temperatures, ranging from approximately $300\ \text{K}$ to $370\ \text{K}$ in very extreme cases \cite{KU2019Predicting}. For biological relevance, we consider a threshold temperature of $300\ \text{K}$, and non-dimensionalize it using representative protein values for length, stiffness, and strain.

We consider hemoglobin as a representative protein due to its well-studied cooperative function, and make our estimates accordingly. The total size of a hemoglobin macromolecule is $\sim 5\ \text{nm}$ \cite{Erickson2009Biol.Proced.Online}. We model proteins using a spring network with 128 nodes; the corresponding equilibrium length of bonds therefore translates to $\sim 1\ \text{nm}$. Recent advances in modeling Young’s moduli of proteins based on molecular interactions allow a distinction between bulk properties and the interface region, estimating the former at $Y\sim10\ \text{GPa}$ \cite{Song2021PLOSONE}. From this, we estimate the typical coarse-grained spring stiffness to be $\sim 10\ \text{N/m}$.

To identify a representative range for the binding-induced source strain $\varepsilon_s$, we use the topological analysis developed in~\cite{rocks2024topological} to identify the source sites in actual allosteric protein structures, and observed the binding-induced deformations reported in the Protein Data Bank~\cite{ProteinBank}. We analyze the distribution of strain between individual atom pairs at the source for 24 different proteins, calculating a median value for the source strain $\varepsilon_s=0.19$, and a 25th–75th percentile range $0.09$ to $0.31$ to reflect inter-protein variability. We further estimated the size of these proteins by measuring the distance between their furthest atoms. The median length is $5.4\ \text{nm}$, and a 25th–75th percentiles are between $4.7$ and $6.6\ \text{nm}$.

A rough estimate of a typical binding strain can also be derived from energy considerations. For hemoglobin, the binding energy is approximately $0.5 \mathrm{eV}$~\cite{Chow1975Studies, Alcantara2007Quantum}. Assuming an interaction length scale $l \sim 1\ \text{\AA}$, the effective interaction stiffness is then $k_I\sim lY=1\ \text{N/m}$. To connect these values with the dimensionless binding energies $E^A$ and $E^B$ ($\sim200$, Fig.~\ref{fig1}C) we assume the binding energy scales as $\sim\varepsilon_s^2$. Under these assumptions, the required source strain is $\varepsilon_s \approx0.2$ in good agreement with our estimates based on protein structures.

Finally, to compute the protein-equivalent $300\ \text{K}$ threshold for Figs. \ref{fig2} (dashed gray line), we use $k_B \cdot300 \mathrm{K} / \left(\varepsilon_s^2 (10 \mathrm{N/m})(1 \mathrm{nm})^2\right)$. To  estimate the 25th-75th percentile range for the temperature, we used the strain and length of the same protein together, and then compounded the results across the different proteins. This yields an estimated range of values, shown as the gray band in Fig.~\ref{fig2}, against which we compare simulation results for $\Theta^*/\varepsilon_s^2$.

Note that the crossover temperatures of our tuned networks do not necessarily lie above the gray band, suggesting that cooperative function might not remain robust at biologically relevant temperatures. Clearly, however, the cooperative function of real proteins \emph{does} remain robust at relevant temperatures. This suggests that it should be possible for tuned networks to have higher crossover temperatures. We show next that indeed it is possible to modify the tuning process to raise the crossover temperature significantly. 

\section{Raising the Crossover Temperature}

From Eq. \ref{crossover condition}, we see that the crossover temperature is set by the competition between the mean strain response of the target in the bound case (which increases with $\Delta$) and the strain fluctuations in the unbound case (which also increases with $\Delta$). Clearly, one can increase the crossover temperature by decreasing fluctuations. This can be accomplished simply by increasing all spring stiffnesses by the same factor. That is not a realistic strategy for proteins, however, where the range of available stiffnesses is limited, and cannot simply be increased across the board. Our aim is therefore to raise the \emph{dimensionless} crossover temperature $\Theta^*$, where the trivial scaling with spring stiffnesses is cancelled out.  

Recall from Fig.~\ref{fig1}E-F that the node displacements from bidirectional allostery project strongly onto a single low-frequency vibrational mode. Eq.~\ref{strain fluctuations} shows that increasing the frequency of that mode would reduce target fluctuations in the unbound state, potentially raising the dimensionless crossover temperature, $\Theta^*$. We show here that an effective strategy for increasing $\Theta^*$ is to tune the free state to have a higher energy while still preserving function. 

Coupled Learning provides a natural framework to tune the network while simultaneously raising the energy of the free state. We introduce a generalized contrast function:
\begin{equation}\label{eq: modified cost function}
\mathcal{\tilde{C}} = (1-r) \left(E^C-E^F\right) - rE^F,
\end{equation}
where the parameter $r$ controls the tradeoff between raising the free state energy ($r=1$) and attaining allosteric function ($r=0$). For symmetric tuning (increased cooperativity), the learning rule becomes:
\begin{equation}\label{eq12: regularization learning rule}
\begin{aligned}
\frac{dk_i}{dt}=-\frac{\alpha}{\eta}\frac{d}{d k_i}\left( \mathcal {\tilde{C}}_{A\rightarrow B} + \mathcal {\tilde{C}}_{B\rightarrow A}\right).
\end{aligned}
\end{equation}

Because it is still important to maintain cooperativity, we want to choose $r$ such that $rE^F \ll E^C - E^F$. We can approximate $E^F \sim \langle k_i l_i^2 \rangle \varepsilon_s ^2$. The energy difference between the free and clamped states depends on the nudge factor $\eta$ and the performance error $\varsigma = |\Delta-\hat{\Delta}|/\Delta$, $\left( E^C - E^F \right) \sim \langle k_i l_i^2 \rangle \varepsilon_s ^2 \eta \varsigma$. We thus should choose $r\ll\eta\varsigma$. 
For $\eta=0.01$, and aiming for solutions with error $\varsigma\sim1\%$, we restrict ourselves to $r \ll 10^{-4}$. 

Figure~\ref{fig2}D depicts aggregated results for the dimensionless crossover temperature $\Theta^*$ for various $r$, tuned for $0.05 \le \Delta \le 0.95$ with a performance error of $|\Delta-\hat{\Delta}|/\Delta \le 0.025$. For all $r$ values, the crossover temperature still vanishes at $\Delta=0$ because $\langle \varepsilon_t \rangle=0$. It also vanishes at $\Delta=1$ because the cooperative mode becomes floppy. Remarkably, as shown in Fig.~\ref{fig2}, we find that for a range of $r$ the networks can preserve their functionality at biologically relevant temperatures. The highest crossover temperature is obtained when $\Delta \approx 0.65$ to $0.7$.

The benefit of increasing $r$ is non-monotonic; $\Theta^*$ increases up to $r^\dagger \sim 5 \times 10{-6}$ and then decreases for higher $r$.  At $r^\dagger$ the springs important for the function have already attained the maximum stiffness allowed. For $r \gtrsim r^\dagger$, less important springs are also stiffened, raising the dimensional
crossover temperature $T^*$ (see \cite{supplemental}) 
but lowering $\Theta^*$.

\section{Melting Temperature}
To maintain allosteric functionality, not only must the dimensional crossover temperature $T^*$ but also the denaturation (melting) temperature $T_M$ must lie above biologically relevant temperatures. 
While $T^*$ is controlled by thermal fluctuations at specific target sites, $T_M$ reflects the loss of overall structural stability due to fluctuations across the entire network. To quantify these, we adopt a generalized Lindemann parameter.

Originally introduced to capture the solid-to-liquid transition, the Lindemann criterion hypothesizes that crystalline solids melt when the ratio $L$ of the root-mean-square 
atomic fluctuations to interatomic spacing exceeds some threshold, $L_M$, where $L_M \sim 0.15$ \cite{Lindemann1984calculation}. This concept has been extended to finite inhomogeneous systems such as proteins by replacing the interatomic spacing with the mean nearest-neighbor distance \cite{Stillinger1990JChemPhys, Zhou1999JMB, Faraggi2018J.Biomol.Struct.Dyn.}. For such systems, melting transitions occur at a threshold $L_M$ in the range $0.14-0.19$, remarkably close to the value for crystalline solids \cite{Zhou1999JMB,Marina2017pnas}.
In our spring system, the Lindemann parameter can be expressed in linear response (\textit{Appendix \ref{apA}}) in terms of the vibrational modes as:
\begin{equation}\label{Lindemann criterion}
L = \sqrt{\frac{\frac{1}{N}\sum_j \frac{k_BT}{m\omega^2_j}}{\langle l_i^2 \rangle}},
\end{equation}
and as before, we exclude the trivial zero modes from the sum over eigenmodes $j$. 
Here, $l_i$ are the equilibrium lengths of bonds. The melting temperature is then
\begin{equation}\label{Melting Temperature}
k_B T_M = L_M^2\frac{\langle l_i^2 \rangle}{\frac{1}{N}\sum_j \frac{1}{m\omega^2_j}}.
\end{equation}

\begin{figure}
\includegraphics[width=.99\linewidth]{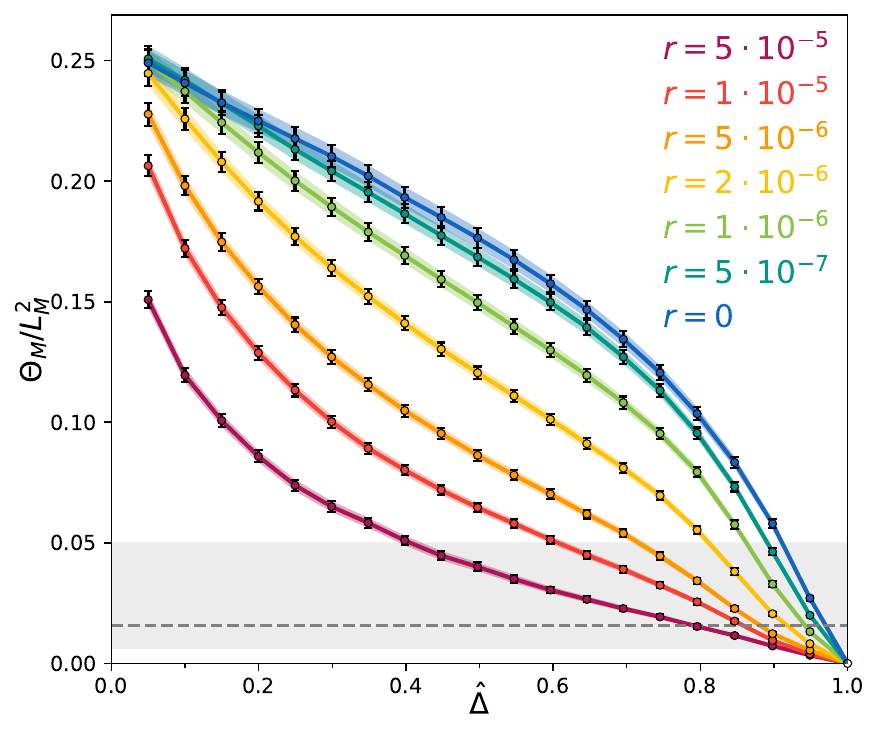}
\centering
\caption{\textbf{Melting Temperature.} The dimensionless melting temperature $\Theta_M$ as a function of attained response ratio $0.05\leq\hat{\Delta}\leq0.95$ at increasing values of the parameter $r$. The colors match the scheme in Fig. \ref{fig2}. 
$\Theta_M=0$ at $\hat{\Delta}$ because the systems become floppy.
Vertical and horizontal bars (too small to see) indicate standard errors. The gray dashed line (shaded region) corresponds to $300\ \text{K}$, obtained by matching our coarse-grained model with median (25th-75th percentile) protein values.
}
\label{fig3}
\end{figure}

Notice that $T_M$ depends on $L_M^2$, similarly to how $T^*$ depends on $\varepsilon_s^2$ (Eq. \ref{crossover condition}). Figure \ref{fig3} illustrates the dimensionless melting $\Theta_M = k_BT_M/\langle k_i l_i^2\rangle$ as functions of $\Delta$ for several values of $r$ (defined in Eq. \ref{eq: modified cost function}). $\Theta_M$ was further scaled by $L_M^2$ to offset the dependence on the Lindemann threshold. 
According to Eq.~\ref{Melting Temperature}, the melting temperature approaches zero when any eigenvalue becomes zero; hence, $\Theta_M=0$ at $\Delta=1$, where the system develops a floppy mode.
Tuning the system for progressively larger $\Delta$ leads to a reduction in the melting temperature, due to softening of the cooperative mode (blue curve in Fig. S25). 
For a given $\Delta$, however, the system stiffens with increasing $r$, resulting in an increasing $T_M$ with $r$ (Fig. S25). The \emph{dimensionless} melting temperature $\Theta_M$, however, consistently decreases with increasing $r$. Thus, $r$ raises the dimensionless crossover temperature $\Theta^*$ but lowers the dimensionless melting temperature $\Theta_M$. The biologically relevant regime of temperatures is again represented by the gray shaded region, assuming Lindemann parameters $L_M$ in the range $0.1$ and $0.2$.

\begin{figure}
\includegraphics[width=.99\linewidth]{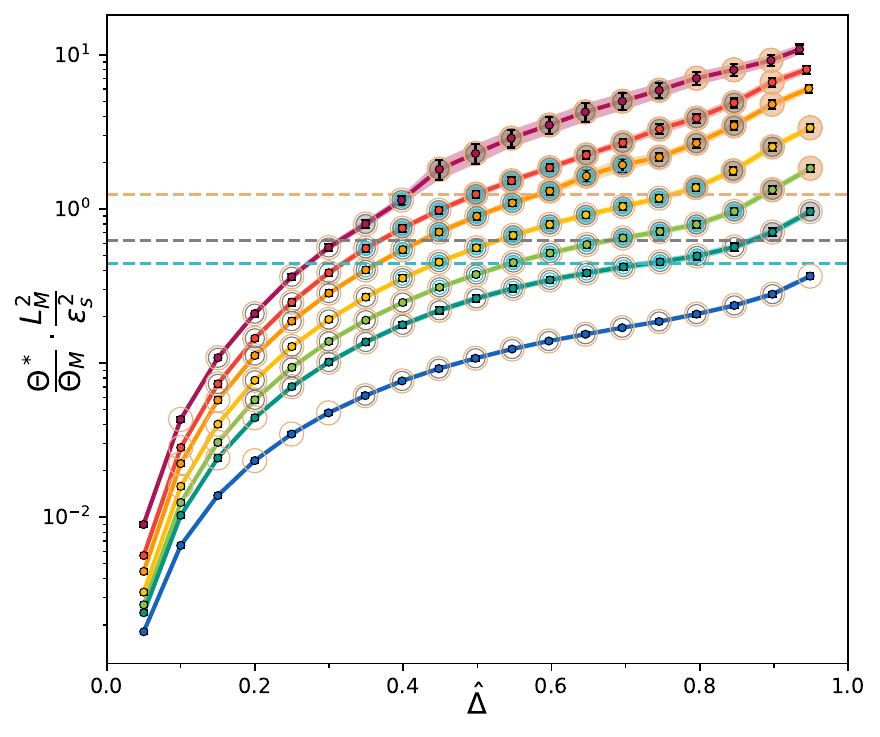}
\centering
\caption{\textbf{Crossover to melting ratio.} The ratio between the crossover temperature and the melting temperature as a function of the attained  response ratio $0.05\leq\hat{\Delta}\leq0.95$, at increasing values of the parameter $r$. 
Orange, gray, and cyan dashed lines indicate where $\Theta^*=\Theta_M$ (functionality is maintained all the way to melting) for choices of $L_M$ and $\varepsilon_s$ corresponding to the 25th, median, and 75th percentile, respectively.
Open circles of the same colors around data points indicate that both temperatures exceed the $300\ \text{K}$ threshold, calculated using the 25th, median, and 75th percentile values. These thresholds correspond to the bottom of the gray band, the gray dashed line, and the top of the gray band in Figs.~\ref{fig2}-\ref{fig3}.
Filled circles indicate that, in addition to both being above the $300\ \text{K}$ threshold, the crossover temperature is also larger than the melting temperature.
}
\label{fig4}
\end{figure}

Clearly, actual proteins with cooperative binding must have values of $T^*$ and $T_M$ that both lie above the biologically relevant range. Figure \ref{fig4} shows the ratio $\Theta^*/\Theta_M$, allowing us to assess if functionality is maintained all the way to the melting point (dashed lines). We find a range of values of $\Delta$ and $r$ where both dimensional temperatures lie above $300\ \text{K}$ (circled points).

\section{Discussion}
We have shown that cooperative binding can be achieved via a symmetric application of allosteric tuning between different sites in elastic networks. Strong strain coupling between the sites leads to higher cooperativity. As demonstrated in previous works using evolutionary algorithms~\cite{Yan2018Principles}, we find that the functional mechanism relies on the development of a low-frequency mode. The existence of such a mode threatens stability of function at higher temperature because it implies large fluctuations at the target.

Our work goes beyond previous studies that tuned mechanical spring networks to develop protein allostery or cooperativity which only considered function at zero temperature ~\cite{Hawkins_McLeish2004,Rocks2017Designing, Yan2017PNAS, Yan2018Principles, Rocks2019Limits, Ravasio2019Mechanics, Rouviere2023Emergence, rocks2024topological}, or studies that focused only on fluctuations, where the temperature is a pre-factor to the free-energy barrier controlling the rate of the binding/unbinding \cite{Cooper1984,Hawkins_McLeish2004, Zhu2011}.
This paper introduces a theoretical framework for how thermal fluctuations affect allostery and cooperativity as a function of temperature. From the theory, we identify a crossover temperature, $T^*$, above which functionality is destroyed by thermal fluctuations. In comparing our results to proteins, we find that $T^*$ can lie below biologically relevant temperatures for networks tuned only for allosteric functionality. To raise $T^*$, we tune using a modified contrast function, which selectively stiffens the mode responsible for cooperative function, to a degree controlled by the parameter $r$. We find that $r$ lowers the melting temperature, but that there is a range of values of $\Delta$ and $r$ for which both the crossover temperature and the melting temperature lie above biologically relevant temperatures.

Of course, proteins were tuned by the evolutionary process, not by the process that we have used to tune our networks. Our work should be considered a proof of principle, showing that it is possible to obtain networks with cooperative function that is robust at biologically relevant temperatures.

\begin{figure}
\includegraphics[width=.99\linewidth]{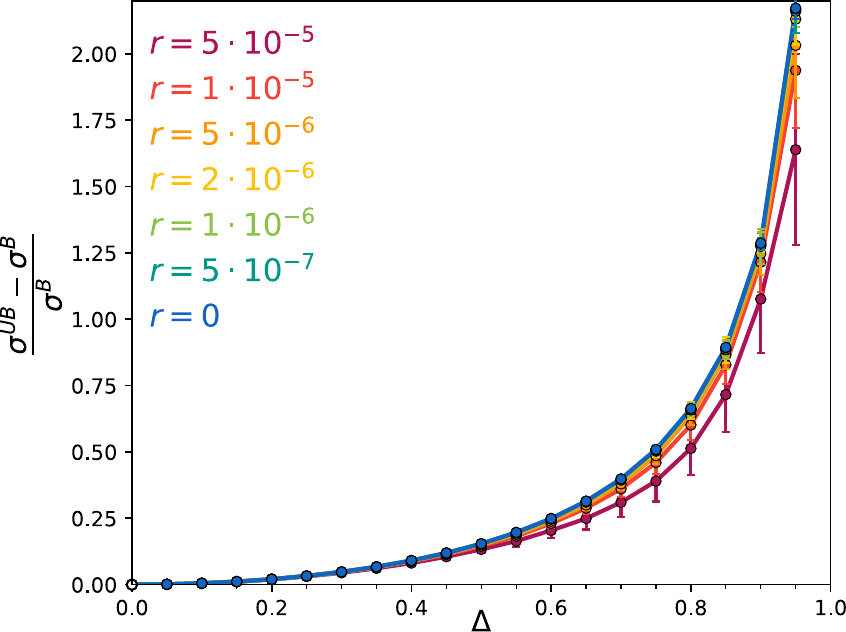}
\centering
\caption{\textbf{Change in fluctuations.} 
The difference between the target's strain fluctuations at the unbound ($\sigma^{UB}_{\varepsilon_t}$) and bound ($\sigma^{B}_{\varepsilon_t}$) states, as a function of the allosteric ratio $\Delta$ at increasing values of the parameter $r$ defined in Eq.~\ref{eq: modified cost function}. The tuning process weights thermal stability more strongly with increasing $r$. As $\Delta$ increases, the fluctuations in the unbound state become dominant. The difference $\sigma^{UB}_{\varepsilon_t}- \sigma^{B}_{\varepsilon_t}$ also
grows with temperature as $\sim \sqrt{T}$.
The results for increasing $r$ values collapse well, suggesting that the coupling between conformational and dynamic allostery is a robust characteristic of the functionality rather than of the tuning process.}
\label{fig5}
\end{figure}

Because we tune in functionality at $T=0$, however, the mechanisms underlying cooperative function in our models are fundamentally energetic in origin, involving strains propagating within the network. In the protein literature, this form of allostery is known as ``conformational allostery". 
In proteins, which necessarily function at nonzero temperature, allostery is a dynamic process \cite{Kern2003, Popovych2006, Tsai2008,Goodey2008}.
The literature points to another form of allostery, called "dynamic allostery", which can occur even in the absence of any conformational change~\cite{Cooper1984, Hawkins_McLeish2004, Liu_Hilser2007, Zhu2011, Wrabl_Hilser2011}. In this nonzero-temperature phenomenon, fluctuations of the target depend on whether the source is bound or not. This is viewed as an entropic effect.
Figure \ref{fig5} shows that in our systems, tuned for conformational allostery at $T=0$, the target fluctuations at $T>0$ differ in the bound and unbound case, consistent with dynamic allostery. As discussed earlier, when the system is trained to exhibit a high $\Delta$, its response to binding is governed by a low-frequency mode. This mode also dominates fluctuations of both the source and target site in the unbound state ($\sigma^{UB}_{\varepsilon_t}$) but does not have a counterpart in the bound state. In that case, fluctuations in the source site are suppressed via the addition of a stiff spring. This suppresses the contribution of the lowest-frequency mode to fluctuations at the target site so that target fluctuations (quantified by $\sigma^{B}_{\varepsilon_t}$) are markedly smaller.
The difference in the fluctuations in the unbound and bound cases increases with $T$ and $\Delta$. Thus, our networks, which are tuned to exhibit conformational allostery, also display the characteristic features required for dynamic allostery, showing that entropic and conformational effects can go hand in hand. This result raises an important direction for future work, to develop simple models of allostery with entropic as well as energetic mechanisms. In particular, models with prestress lead to frustration and a rugged energy landscape \cite{Rouviere2023Emergence}, suggesting that even richer possibilities for such mechanisms can arise at nonzero temperature and prestress.

\begin{acknowledgments}
We thank Menachem Stern for insightful discussions. This work was supported by DOE Basic Energy Sciences through grant DE-SC0020963 (AJL, BP, JA) and DE-SC0020972 (SRN) and the Simons Foundation through Investigator grant \#327939 to AJL.  AJL is also grateful for the hospitality of the Aspen Center for Physics (NSF grant PHY-2210452) and the Santa Fe Institute.
\end{acknowledgments}

\appendix
\section{Linear Response Theory}\label{apA}

The node displacements of a network are a sum of two contributions: an athermal part $\vec{u}$ resulting from stresses on the bonds due to the imposed strain at the source, and a thermal contribution $\vec{v}$ stemming from thermal fluctuations. For a three-dimensional network of $N$ nodes and $N_b$ bonds, $\vec{u}$ and $\vec{v}$ contain $3N$ entries, one for each particle's degree of freedom. In the linear regime, $\vec{u}$ is derived using the Hessian and rigidity matrices, $\mathbf{H}$ and $\mathbf{R}$: 
\begin{equation}\label{athermal displacements}
\vec{u} = \mathbf{-H^{-1}}\vec{f} = \mathbf{-H^{-1}R^T}\vec{\tau}  = \mathbf{H^{-1}R^{T}KL}\vec{\varepsilon_{in}},
\end{equation}
where $\vec{f}$ is the force applied to the nodes and $\vec{\tau}$ is the $N_b$-long vector of bond stresses, which in turn can be written using the $N_b\times N_b$ diagonal stiffness and rest length matrices $\mathbf{K}$ and $\mathbf{L}$, and the $N_b$-long input strain vector $\vec{\varepsilon_{in}}$, which is $\varepsilon_s$ at the source and zero everywhere else. The thermal contribution $\vec{v}$ can be expressed using the projection on the normal modes of the system $\{ \vec{e_j} \}$,
\begin{equation}\label{thermal displacements}
\vec{v} = \sum\limits_{j=7}^{3N} a_j \vec{e_j}.
\end{equation}
where we have excluded the 6 zero modes corresponding to global rotations and translations. The random coefficients $a_j$ follow a zero-mean normal distribution stemming from equipartition,
\begin{equation}\label{coefficients}
a_j \sim \mathcal{N}(\mu=0, \sigma^2 = \frac{k_BT}{m\omega^2_j}).
\end{equation}
where $m\omega^2_j$ is the eigenvalue corresponding to mode $\vec{e_j}$.

We then convert node displacements to strain using the rigidity matrix and the inverse of the rest length matrix. Combining the two contributions, the target strain is:
\begin{equation}\label{target strain total}
\varepsilon_t = \frac{l_s \varepsilon_s}{l_t} \left[\mathbf{RH^{-1}R^T K}\right]_{t,s}+ \frac{1}{l_t} \sum\limits_{j=7}^{3N} a_j \ \vec{R}_t \cdot \vec{e_j},
\end{equation}
where $l_t$ and $l_s$ are the rest length of the target and source sites, $\vec{R}_t$ is the $t^{\mathrm{th}}$-row of matrix $\mathbf{R}$, and subscripts $t,s$ indicate the entry of the bracketed matrix that corresponds to the target and source bond indices, respectively. 

\section{Fluctuations in Linear Response Theory} \label{apB}
Eq. \ref{target strain total} describes the target strain as a sum of a mechanical and a thermal component.
Written this way, we see that the resulting strain fluctuations stem from the second component, and using Eq. \ref{coefficients}, can be expressed as follows:
\begin{equation}\label{strain fluctuations 2}
\begin{split}
\sigma^2_{\varepsilon_t} & = \mathbb{E}[\left(\varepsilon_t\right)^2] -  \mathbb{E}[\varepsilon_t]^2   \\ & = \frac{1}{l^2_t} \sum\limits_{j=7}^{3N} \frac{k_BT}{m\omega^2_j} \left( \vec{R}_t \cdot \vec{e_j} \right)^2 \equiv k_BT \tilde{\sigma}^2_{\varepsilon_t}.
\end{split}
\end{equation}
where $\tilde{\sigma}^2_{\varepsilon_t}$ is the combined mechanical pre-factor from all the modes. 
Written this way, we find that the contribution from each mode results from two factors; the first is how much the mode is excited by energy,  which is inversely proportional to the mode's frequency squared; the second 
is the mode's specificity to the bond, namely, the extent to which exciting this mode results in changing the length of this specific bond. 

In the unbound case, where no external strain is applied at the source ($\varepsilon_s=0$), the first term in Eq. \ref{target strain total} vanishes and $\langle \varepsilon_t \rangle=0$ because the coefficients $a_j$ are drawn from a distribution with zero mean (Eq.~\ref{coefficients}). The target strain thus fluctuates about zero with the ``unbound" standard deviation $\sigma^{UB}_{\varepsilon_t}$ given by Eq. \ref{strain fluctuations 2}. For systems tuned for high $\Delta$ values, $\sigma^{UB}_{\varepsilon_t}$ is dominated by the contribution of the allosteric mode. This low-frequency mode is easily excited and closely aligned with the free-state conformation, in which the target (and source) site experiences high strain. As a result, it exhibits high specificity for the target site.

In the bound case, $\left< \varepsilon_t \right>\neq0$ and approaches $\Delta\varepsilon_s$ at the end of tuning. In this case, the fluctuations $\sigma^{B}_{\varepsilon_t}$ are derived by considering a modified system where a stiff spring (two orders of magnitude stiffer than the stiffest spring of the unbound system) is added to the source bond to fix it close to the desired strain. If $l$ is the length of the source site, then the rest length of this bind-enforcing spring is set to be $\left(1+\varepsilon_s\right)l$. The modified system does not support the allosteric mode, and therefore, no corresponding increase in $\sigma^{B}_{\varepsilon_t}$ is recorded in systems tuned for high $\Delta$.

\bibliographystyle{apsrev4-2}

%

\end{document}